\documentclass[aps,showpacs,twocolumn,twoside,prl]{revtex4-1}

\usepackage{graphicx}  
\usepackage{adjustbox}
\usepackage[version=4]{mhchem}
\usepackage[percent]{overpic}  
\usepackage{mhchem}  
\usepackage{epstopdf}
\usepackage{amsmath}
\usepackage{amssymb}
\usepackage{bm}
\usepackage{xcolor,xspace}
\usepackage{hyperref}
\hypersetup{colorlinks=True,urlcolor=blue,linkcolor=blue,
citecolor=blue,filecolor=black}
\usepackage{braket}
\usepackage[capitalise]{cleveref}  

\definecolor{CornflowerBlue}{rgb}{0.39, 0.58, 0.93}
\definecolor{Salmon}{rgb}{0.98, 0.50, 0.45}
\definecolor{MediumSeaGreen}{rgb}{0.24, 0.70, 0.44}
\definecolor{Goldenrod}{rgb}{0.95, 0.75, 0.15}
\definecolor{MediumPurple}{rgb}{0.57, 0.44, 0.73}
\definecolor{FireBrick}{rgb}{0.70, 0.13, 0.13}
\definecolor{Turquoise}{rgb}{0.25, 0.88, 0.82}
\definecolor{Chartreuse}{rgb}{0.498, 1, 0}
\definecolor{SlateBlue}{rgb}{0.416, 0.353, 0.804}

\begin{document}
\title{Competing shape evolution, crossing configurations 
and single particle levels in nuclei}
\author{N.~Gavrielov}\email{noam.gavrielov@mail.huji.ac.il}
\affiliation{GANIL, CEA/DSM–CNRS/IN2P3, Bd Henri Becquerel, 
BP 55027, F-14076 Caen Cedex 5, France}
\affiliation{Racah Institute of Physics, The Hebrew 
University, Jerusalem 91904, Israel}

\date{\today}  

\begin{abstract}
The evolution of shape in the even-even zirconium (Zr) 
isotopes has been the subject of study for many years. 
However, the odd-mass isotopes have not been investigated 
as extensively due to limited experimental accessibility 
and computational challenges. 
This work, employing the interacting boson-fermion model 
with configuration mixing, examines the effect of rapid 
shape evolution and normal-intruder configuration crossing 
--- both identified as quantum phase transitions --- 
alongside evolution in single particle energies, on the 
positive-parity spectrum of odd-mass $^\text{93--103}$Zr 
isotopes.
Calculated energy levels, magnetic moments, $B(E2)$ values, 
and quadrupole moments are compared to experimental data, 
showing good agreement. 
The special case of $^{99}$Zr, which lies near the critical 
point of both quantum phase transitions, is also addressed, 
offering a new interpretation to the $7/2^+_1$ isomeric 
state and the occurrence of the type II shell evolution, in 
light of recent debates.
\end{abstract}

\maketitle

The study of atomic nuclei reveals a diverse array of 
phenomena, from single particle excitations to collective 
motions and shape coexistence. These phenomena are crucial 
for understanding the evolution of nuclear structure across 
different isotopes.
In particular, the evolution of effective single particle 
energies (ESPE) and the resulting formation of collective 
bands \cite{Otsuka2019} provide insight into the interplay 
between nucleon interactions and nuclear shape changes.
 
Often, the evolution in shapes discloses a quantum phase 
transition (QPT) \cite{Cejnar2010} that is apparent in the 
spectrum, where the shape is the phase that changes 
as one varies the control parameter, the number of nucleons 
($A$). Furthermore, coexisting shapes have been studied 
extensively, and are shown to occur due to 
particle-hole excitations across multiple shell model 
configuration \cite{Heyde2011}. 
In certain cases, these configurations can cross, leading 
to a different kind of QPT \cite{Frank2006} with the phase 
being the dominant shell model configuration in the ground 
state wave function.

The scenario of configuration crossing QPT is rather unique 
and was identified in the even-even zirconium (Zr) 
isotopes, attributed to proton-neutron interactions within 
spin-orbit partner orbitals \cite{Federman1979}, i.e. 
orbitals with $j_\pi=\ell \pm 1/2$ and $j_\nu = \ell \mp 
1/2$. Subsequent studies found that such spin-orbit 
partners can cause a type II shell evolution of the EPSEs, 
influencing the QPT \cite{Togashi2016}.

The effect of single particle degrees of freedom is even 
more accentuated in odd-mass nuclei, where the nucleus can 
be viewed as an even-even core coupled to an extra 
fermion. Nevertheless, theoretical investigations of 
odd-mass nuclei, remain limited. 
This gap is particularly evident in the study of odd-Zr 
isotopes, where previous works mentioned the need for 
incorporating configuration mixing 
\cite{Brant1998,Lhersonneau2005,Rodriguez-Guzman2011,
Spagnoletti2019a,Boulay2020,Nomura2020,Pfeil2023}.

Recent studies have demonstrated that the configuration 
crossing QPT in the even-even Zr isotopes is intertwined 
with a shape-evolution QPT, leading to a scenario termed 
intertwined QPTs (IQPTs) 
\cite{Gavrielov2019, Gavrielov2020,Gavrielov2022}. 
Subsequently, using the new interacting boson-fermion model 
with configuration mixing (IBFM-CM) \cite{Gavrielov2022c}, 
the spectrum of odd-mass niobium (Nb) isotopes also 
revealed the occurrence of IQPTs 
\cite{Gavrielov2022c,Gavrielov2023a}. This underscored the 
necessity of incorporating multiple configurations to fully 
understand the structure of odd-mass nuclei.

The odd $^\text{93--103}$Zr isotopes exhibit a rapid change 
in ground state total angular momentum: $5/2^+$, $5/2^+$, 
$1/2^+$, $1/2^+$, $3/2^+$, and $5/2^-$. This change, caused
by neutron orbit occupations that shift the Fermi level, is 
key to understanding the behavior of these isotopes. In 
heavier Zr isotopes, the evolving single quasi-particle 
energies (SqPEs) affect the spectrum. The SqPEs evolution, 
driven also by the configuration crossing QPT and shape 
evolution QPT within the even-even Zr core 
\cite{Gavrielov2022}, contrasts with the Nb isotopes study
\cite{Gavrielov2022c,Gavrielov2023a}, where the Fermi level 
remains unchanged.

It is therefore the goal of this paper to unravel the 
effect on the spectrum and the interplay between the triad 
evolution of shapes, configurations, and single 
quasi-particle energies in the odd $^\text{93--103}$Zr 
isotopes.

The IBFM-CM is employed to study odd-$A$ nuclei 
\cite{Gavrielov2022c, Gavrielov2023a}. In the IBFM 
\cite{IachelloVanIsackerBook}, a boson core represents the 
adjacent even-even nuclei and is coupled to a fermion 
($a$). 
The core is described using the interacting boson 
model with configuration mixing 
\cite{IachelloArimaBook,Duval1981,Duval1982}, where $N_b$ 
monopole ($s$) and quadrupole ($d$) bosons represent 0p-0h 
excitations and 2p-2h core excitations with $N_b+2$ bosons. 
The total Hamiltonian can be written in the form of
\onecolumngrid
\begin{equation}\label{eq:ham-bf}
\hat H
= \hat H_{\rm b} + \hat H_{\rm f} + \hat H_{\rm bf} 
=
\left [
\begin{array}{cc}
\hat H_\text{b}^{\rm A}(\eta^{\rm(A)}) & \hat 
V_\text{b}(\omega) 
\\ 
\hat V_\text{b}(\omega)  & \hat H_\text{b}^{\rm 
B}(\eta^{\rm(B)})
\end{array}
\right] 
+ 
\left [
\begin{array}{cc}
\sum_j \epsilon_j \hat n_j &  0\\ 
	0  & \sum_j \epsilon_j \hat n_j
\end{array}
\right] 
+
\left [
\begin{array}{cc}
  \hat H^{\rm A}_\text{bf}(\zeta^{(\rm A)}) & 0 \\
  0 & \hat H^{\rm B}_\text{bf}(\zeta^{(\rm B)})
\end{array}
\right],
\end{equation}
\twocolumngrid

In \cref{eq:ham-bf}, $\hat H_{\rm b}$ represents the boson 
Hamiltonian. Here, $\hat H_\text{b}^{\rm A}(\eta^{\rm(A)})$ 
describes the normal A~configuration with $N_b$ bosons,
while $\hat H_\text{b}^{\rm A}(\eta^{\rm(A)})$ describes 
the intruder B~configuration with $N_b+2$ bosons.
Here, a typical Hamiltonian for both configurations is used
\begin{equation}\label{eq:H_b_consis}
\hat H_{\rm b}^i = \epsilon^{(i)}_d \hat n_d + 
\kappa^{(i)} \hat Q_\chi \cdot \hat Q_\chi + \kappa^{\prime 
(i)} \hat L\cdot \hat L + \delta_{i,{\rm B}}\Delta_\text{b},
\end{equation}
with $i=\text{A, B}$. Interactions in \cref{eq:H_b_consis} 
include the \mbox{$d$-bosons} number operator $\hat n_d = 
\sqrt{5}(d^\dagger \tilde d)^{(0)}$, the rotational term 
$\hat L = \sqrt{10}(d^\dagger \tilde d)^{(0)}$, and the 
quadrupole operator 
\begin{equation}\label{eq:quad_op}
\hat Q_\chi = (d^\dag s+s^\dag \tilde d)^{(2)} + \chi 
(d^\dag \tilde d)^{(2)}~,
\end{equation}
where $s$ and $d$ are spherical tensors, coupled in a 
tensor product and $\tilde d_\mu  = (-)^\mu d_{-\mu}$. 
The parameter $\Delta_\text{b}$ represents the energy 
offset between configurations A and B, and the mixing term 
$\hat V_{\rm b}$ is given by
\begin{equation}\label{eq:mixing_int}
\hat V_{\rm b} = \omega 
[(d^\dag d^\dag)^{(0)} \!+\! (s^\dag)^2] +\text{H.c.}~,
\end{equation} 
where H.c. stands for Hermitian conjugate.

For $\hat H_\text{f}$, the fermion Hamiltonian, $\hat n_j = 
\sum_\mu a^\dagger_{j\mu}a_{j\mu}$ is the fermion number 
operator, with $\epsilon_j$ as single particle energies for 
each orbit and $j$ the angular momentum of the orbital. 

The Bose-Fermi Hamiltonian $\hat H_{\rm bf}$ consists of 
monopole, quadrupole, and exchange terms with respective 
strengths $A^{(i)}_0, \Gamma^{(i)}_0$, and 
$\Lambda^{(i)}_0$ ($i = \text{A,~B}$), and occupation 
probabilities $(u_j,v_j)$ following the microscopic 
interpretation of the IBFM \cite{Gavrielov2023a}.

After diagonalization, the resulting wave functions are of 
the form
\begin{multline}\label{eq:wf-bf}
\ket{\Psi;J} =
\sum_{\alpha,L,j}C^{(N,J)}_{\alpha,L,j}
\ket{\Psi_{\rm A};[N_b],\alpha,L,j;J} \\
+ \sum_{\alpha,L_,j}C^{(N+2,J)}_{\alpha,L,j}
\ket{\Psi_{\rm B};[N_b+2],\alpha,L,j;J}~,
\end{multline}
Here, $N_b$ is the boson number, $\alpha$ the quantum 
numbers from the boson Hamiltonian, $L$ ($j$) the boson 
(fermion) angular momentum, and $J$ the total angular 
momentum.
The occupation probabilities for the boson number $N_b$, 
the $n_d$ boson number, and the single particle orbitals of 
the wave function can be determined from
\begin{subequations}
\begin{align}
\label{eq:prob_ibfm_bos}
v^2_{(N_i,J)} & =
\sum_{n_d}v^2_{n_d;(N_i,J)} = \sum_{j}v^2_{j;(N_i,J)}~, 
\quad i = {\rm A,B}\\
\label{eq:prob_ibfm_spe}
v^2_{(j;N_i,J)} & = 
\sum_{\alpha,L}|C^{(N_i,J)}_{\alpha,L,j}|^2~,\\
\label{eq:prob_ibfm_nd}
v^2_{(n_d;N_i,J)} & = 
\sum_{\tau,n_\Delta,j,L}|C^{(N_i,J)}_{n_d, \tau, 
n_\Delta,L,j}|^2~.
\end{align}
\end{subequations}
For \cref{eq:prob_ibfm_bos}, $v^2_{(N_{\rm A},J)} + 
v^2_{(N_{\rm B},J)} = 1$, where $N_\text{A} = N_b, 
N_\text{B} = N_b + 2$. It provides the occupation 
probability of $N_b$, indicating the proportion 
of normal ($v^2_{(N_{\rm A},J)}$) and intruder 
($v^2_{(N_{\rm B},J)}$) configurations. 
\cref{eq:prob_ibfm_spe} represents the occupation 
probability of the single particle orbitals, $j$.
\cref{eq:prob_ibfm_nd} represents the occupation 
probability of $n_d$, which indicates the degree of 
deformation. A large $v^2_{(n_d;N_i,J)}$ suggests dominant 
U(5) boson 
symmetry, equivalent to a spherical shape (phase) in the 
geometrical interpretation of the IBM 
\cite{IachelloArimaBook}. If the wave function is 
distributed among several $n_d$ values, it indicates 
deformation. Thus, using 
\cref{eq:prob_ibfm_bos,eq:prob_ibfm_spe,eq:prob_ibfm_nd} one
can analyze the evolution of occupation in the 
configurational, single particle and deformation content of 
the wave functions.

To analyze electromagnetic transitions, the operator with 
multipolarity $L$ for transitions of type $\sigma$ is 
define as
\begin{equation}\label{eq:TsigL}
\hat{T}(\sigma L) =
\hat{T}_{\rm b}(\sigma L) + \hat T_{\rm f}(\sigma L)~,
\end{equation}
with boson $\hat{T}_{\rm b}(\sigma L)$ and fermion $\hat 
T_{\rm f}(\sigma L)$ parts. In this work, we concentrate on 
$E2$ and $M1$ transitions
\begin{subequations}\label{eq:Tem1}
\begin{align}
\hat{T}_{\rm b}(E2) & =
e^{(\rm A)}\hat Q^{(N)}_{\chi} + e^{(\rm B)}\hat 
Q^{(N+2)}_{\chi}~,\label{eq:te2_b}\\
\hat T_{\rm f}(E2) & = 
\sum_{jj^\prime}f^{(2)}_{jj^\prime}[a^\dagger_j \times 
\tilde a_{j^\prime}]^{(2)},\label{eq:te2_f}\\
\hat T_{\rm b}(M1) & =
\sum_i \sqrt{\frac{3}{4\pi}}g^{(i)} \hat L^{(N_i)} 
\nonumber\\
& \qquad\qquad + \tilde g^{(i)}[\hat Q^{(N_i)}_{\chi} 
\times \hat 
L^{(N_i)}]^{(1)}~.\label{eq:tm1_b}\\
\hat T_{\rm f}(M1) & = 
\sum_{jj^\prime}f^{(1)}_{jj^\prime}[a^\dagger_j \times 
\tilde a_{j^\prime}]^{(1)}~.\label{eq:tm1_f}
\end{align}
\end{subequations}

For the $E2$ transitions, \cref{eq:te2_b} defines 
the boson effective charges for configurations A and B as
$e^{(\rm A)}$ and $e^{(\rm B)}$, respectively. In 
\cref{eq:te2_f}, the fermion part is given by 
$f^{(2)}_{jj^\prime} = 
-\frac{e_f}{\sqrt{5}}\braket{j||Y^{(2)}_{lm}||j^\prime}$, 
where $e_f$ is the fermion effective charge.

For the $M1$ transitions, \cref{eq:tm1_b} specifies 
the free value of the neutron spin $g$-factor as $g_s = 
-3.8263~\mu_N$, which is quenched by 30\%, while the 
orbital angular momentum $g$-factor is $g_l = 0~\mu_N$.

The $\ce{_{40}^AZr}$ isotopes with mass number 
\mbox{$A = \text{93--103}$} are described by coupling a 
neutron to their respective even-even $\ce{_{40}^AZr}$ 
cores with $A = \text{92--102}$. 
The parameters of $\hat H_{\rm b}$ in \cref{eq:ham-bf} and 
the boson numbers are thus taken from a previous work 
\cite{Gavrielov2022}, with the sign one $\chi$ in 
\cref{eq:quad_op} modified for $^\text{92--96}$Zr, as 
recent measurements of spectroscopic quadrupole moments of 
$^{94}$Zr have shown to be positive \cite{Marchini2024}.

In the odd $\ce{_{40}^AZr}$ isotopes, the valence neutrons 
occupy different orbits within the $Z\!=\!\text{50--82}$ 
shell. Talmi~\cite{Talmi1962} showed that neutrons in 
$^\text{90--96}$Zr primarily reside in the $\nu 1d_{5/2}$ 
orbital, with $^{96}$Zr representing 
a subshell closure. Therefore, for $^\text{93--95}$Zr, an 
occupation of $n/(2j + 1) = 3/6$ and $5/6$, 
respectively, is assumed, where $n$ is the number of 
valence neutrons.
For $^\text{97--103}$Zr, a BCS calculation is performed 
\cite{Gavrielov2023a} to evaluate the SqPEs ($\epsilon_j$) 
and their occupations ($v^2_j$). As more neutrons are 
added, the changing Fermi energy affects these evaluations. 
According to the microscopic interpretation of the IBFM 
\cite{Scholten1985}, the $\epsilon_j$ are required for 
$\hat H_\text{f}$, and $v^2_j$ impact $\hat H_\text{bf}$ of 
\cref{eq:ham-bf}. The neutrons are distributed among the 
$\nu 2s_{1/2}, \nu 1d_{3/2}, \nu 0g_{7/2}$, and $\nu 
0h_{11/2}$ orbitals, with the $\nu d_{5/2}$ considered 
closed, as done in a previous IBFM calculation 
\cite{Lhersonneau1990}. 
For the BCS calculation, the single particle energies of 
$^{91}$Zr are used \cite{Barea2009}, along with a pairing 
gap of $\Delta_\text{F} = 0.9$~MeV, calculated as the 
average of $\Delta_\text{F} = \frac{1}{4}(B(N-2,Z) - 
3B(N-1,Z) + 3B(N,Z) - B(N+1,Z)$ \cite{BohrMott-II}.

\begin{figure*}[tb!]
\includegraphics[width=1\linewidth]{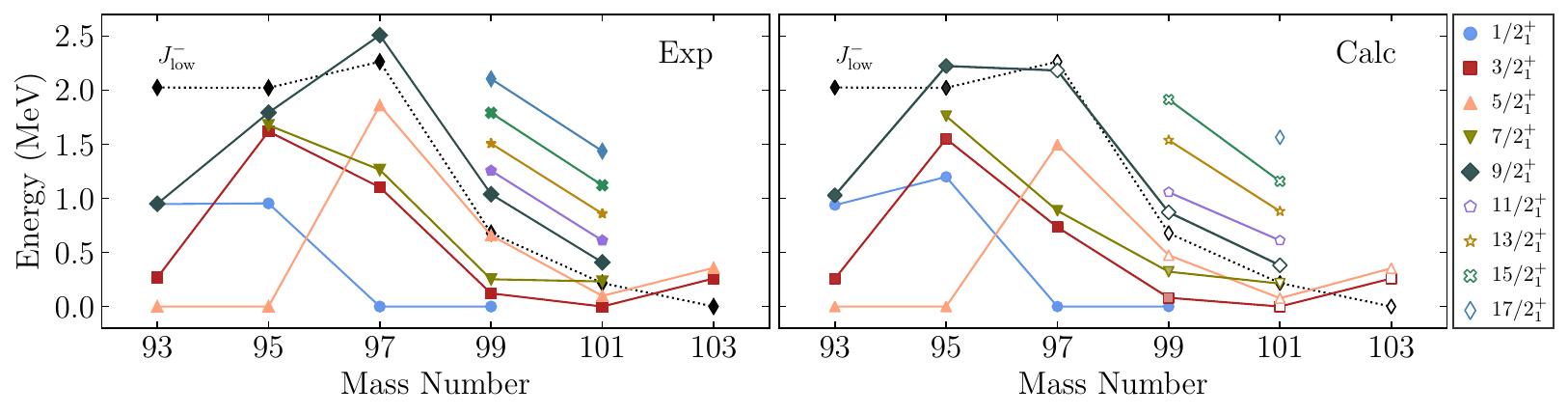}
\caption{Comparison between experimental (left panel) and 
calculated (right panel) lowest-energy levels in odd Zr 
isotopes. On the right panel, the symbols filling indicates 
the dominance of the normal A~configuration (filled) or 
intruder B~configuration (empty) with assignments based on 
\cref{eq:prob_ibfm_bos}. For example, the $3/2^+_1$ is 
normal A~configuration for $A=\text{93--97}$, mixed for 
$A=99$, and intruder B~configuration for 
$A=\text{101--103}$.\label{fig:energies}}
\end{figure*}

\begin{figure}[tb!]
\centering
\begin{overpic}[width=1\linewidth]{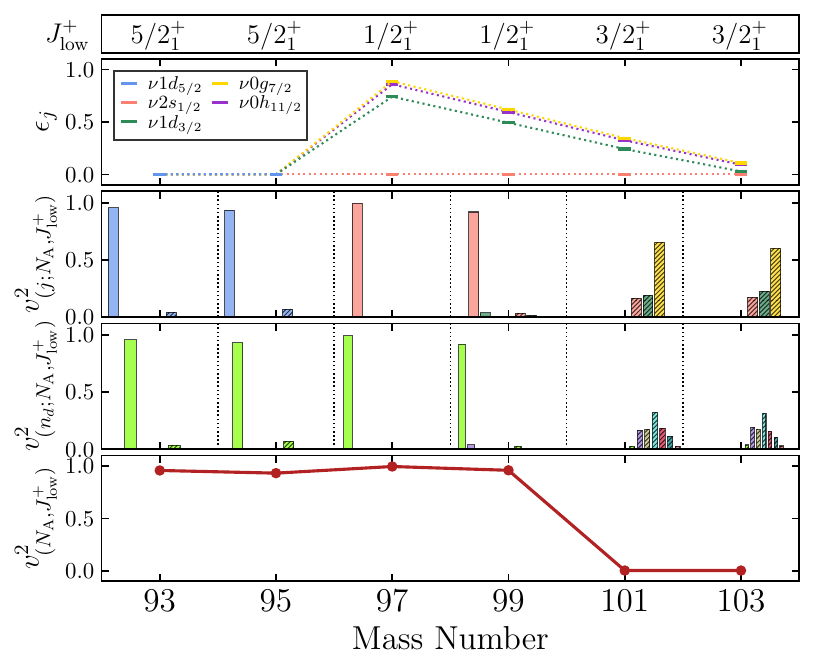}
\put (90,70) {\large (a)}
\put (90,54) {\large (b)}
\put (90,37) {\large (c)}
\put (90,21) {\large (d)}
\end{overpic}
\caption{Evolution of:
(a) Single quasi-particle energies, $\epsilon_j$;
(b) Single quasi-particle occupation probabilities of the 
lowest positive-parity state, $J^+_\text{low}$, 
\cref{eq:prob_ibfm_spe}. Each color indicates a different 
orbital: \textcolor{CornflowerBlue}{\rule{0.5cm}{0.2cm}} 
($d_{5/2}$) \textcolor{Salmon}{\rule{0.5cm}{0.2cm}} 
($s_{1/2}$) \textcolor{MediumSeaGreen}{\rule{0.5cm}{0.2cm}} 
($d_{3/2}$) \textcolor{Goldenrod}{\rule{0.5cm}{0.2cm}} 
($g_{7/2}$). Note the transition from a singular 
single quasi-particle occupation for $^\text{93--99}$Zr to 
a distribution of them for $^\text{101--103}$Zr;
(c) $n_d$ occupation probability of the lowest 
positive-parity state, $J^+_\text{low}$, 
\cref{eq:prob_ibfm_nd}, depicting the different possible 
normal ($n_d=0\ldots,N_b$) and intruder 
($n_d=0\ldots,N_b+2$) $n_d$ values. Each color 
represents a different $n_d$ number from $n_d=0$ 
(\textcolor{Chartreuse}{\rule{0.5cm}{0.2cm}}) to 
$n_d=8$ (\textcolor{SlateBlue}{\rule{0.5cm}{0.2cm}}). 
Note the transition from a single dominant bar 
(representing a spherical state) for $^\text{93--99}$Zr to 
a distribution of them (representing a deformed state) for 
$^\text{101--103}$Zr. In both (b) and (c), bars from the 
left (right) of the mass number correspond to normal 
(intruder) occupations, with intruder bars filled 
with diagonal lines;
(d) the amount of normal occupation in the 
wave function ,\cref{eq:prob_ibfm_bos}. Note the transition 
from normal in $^\text{93--99}$Zr to intruder in 
$^\text{101--103}$Zr. The values of $J^+_\text{low}$ are 
indicated in the uppermost panel.
\label{fig:spe_nd_boson}}
\end{figure}
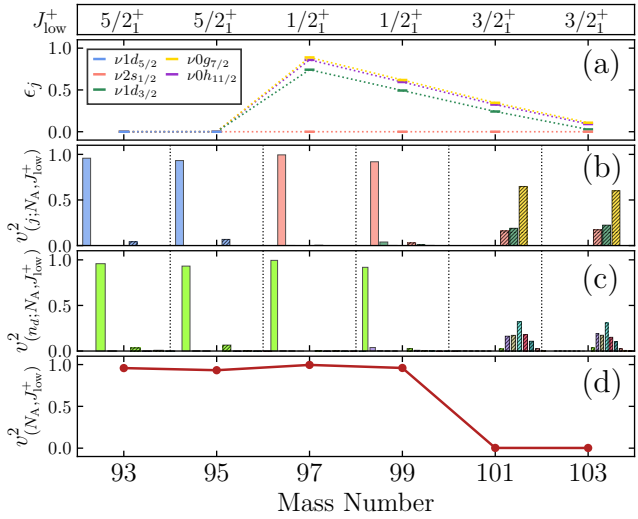

\begin{table}[t]
\begin{center}
\caption{\label{tab:parameters}
\small
Single quasi-particle occupations, the Fermi energy 
$\lambda_F$, and parameters in MeV of the boson-fermion 
interactions, $\hat H^{(i)}_{\rm bf}$ of \cref{eq:ham-bf}, 
obtained from a fit assuming $A^{(i)}_0\!=\!A_0$, 
$\Gamma^{(i)}_0\!=\!\Gamma_0$ and 
$\Lambda^{(i)}_0\!=\!\Lambda_0$ for $i = \text{A and B}$. 
Beneath each isotope the number of valence neutrons used in 
the shell model ($^{93,95}$Zr) and BCS 
($^\text{97--103}$Zr) evaluations is listed.}
\begin{tabular}{ccccccc}
\hline
Isotope & $^{93}$Zr& $^{95}$Zr& $^{97}$Zr& 
				   $^{99}$Zr& $^{101}$Zr& $^{103}$Zr \\
\hline
\# particles & 3 & 5 & 1 & 3 & 5 & 7 \\
\hline
$v^2_{\nu1d_{5/2}}$ & $3/6$  & $5/6$ & 0 & 0 & 0 & 0 \\
$v^2_{\nu2s_{1/2}}$ & 0 & 0 & 0.085 & 0.351 & 0.567 & 
0.697 \\
$v^2_{\nu1d_{3/2}}$ & 0 & 0 & 0.037 & 0.110 & 0.189 & 
0.275 \\
$v^2_{\nu0g_{7/2}}$ & 0 & 0 & 0.033 & 0.091 & 0.152 & 
0.219 \\
$v^2_{\nu0h_{11/2}}$& 0 & 0 & 0.034 & 0.095 & 0.158 & 
0.229 \\
\hline
$\lambda_F$ & 0 & 0 & $-1.339$ & $-0.281$ & 0.122 & 
0.385 \\
\hline\hline
$A_0$		& $-0.252$ & 0.983 & $-0.367$ & 
$-0.150$ & $-0.146$ & 0.133 \\
$\Gamma_0$	& 0.037    & 0.086 & 0.059    & 
0.455    & 0.464    & 0.757 \\
$\Lambda_0$	& 1.191    & 1.063 & 1.121    & 
1.910    & 1.910    & 1.824 \\
\hline 
\end{tabular}
\end{center}
\end{table}
\cref{tab:parameters} presents the resulting $v^2_j$ 
and the strengths $(A_0,\Gamma_0,\Lambda_0)$ for the 
positive-parity states. The strengths for the 
negative-parity states will be addressed in a subsequent 
publication.
The exchange strength $\Lambda_0$ increases gradually along 
the isotopic chain. The quadrupole strength $\Gamma_0$ 
remains relatively constant for $^\text{93--97}$Zr and 
$^\text{99--103}$Zr, suggesting an increase in deformation 
between these regions. The monopole strength $A_0$ 
decreases from $^{93}$Zr to $^{101}$Zr, with exceptions at 
$^{95}$Zr and $^{103}$Zr, possibly due to the lack of 
experimental data.
The occupation probabilities increase with the number of 
valence neutrons, which in turn influences the SqPEs. The 
evolution in SqPEs, shown in 
\cref{fig:spe_nd_boson}(a), depicts a significant lowering 
of the $\nu0h_{11/2}, \nu0g_{7/2}$ and $\nu1d_{3/2}$ 
orbitals from $^{97}$Zr to $^{103}$Zr.

The trends in strengths and SqPEs are reflected in 
the spectrum of $^\text{93--103}$Zr isotopes, shown in 
\cref{fig:energies}, illustrating the energies of selected 
states. The calculated panel, showing good agreement with 
the experimental data, depicts the normal-intruder 
mixing, where filled (empty) symbols represent normal 
(intruder) states, as defined by \cref{eq:prob_ibfm_bos}.
For the lowest positive-parity state, the total angular 
momentum undergoes an intricate evolution: $5/2^+$ for 
$^{93,95}$Zr, $1/2^+$ for $^{97,99}$Zr, and $3/2^+$ for 
$^{101,103}$Zr. 
This evolution results from a combination of deformation, 
configuration mixing, and single  quasi-particle occupation.
For $^{93,95}$Zr, the $5/2^+_1$ state results from the 
$\nu1d{5/2}$ orbital weakly coupled to the respective 
$^{92,94}$Zr cores, particularly the spherical 
normal $0^+_1$. 
A similar situation occurs for $^{97,99}$Zr, where the 
$1/2^+_1$ state is associated with coupling the 
$\nu2s_{1/2}$ orbital to the $^{96,98}$Zr cores. 
For $^{101,103}$Zr, the $3/2^+_1$ state results from 
closely spaced orbitals, ontop of the onset of deformation 
and configuration crossing, where all the lowest lying 
states are intruder.

In \cref{fig:energies} the lowest negative-parity level, 
$J^-_\text{low}$, and its configuration content is also 
shown. For $^{93,95}$Zr it is $11/2^-$ normal,  for 
$^{97}$Zr it is $11/2^-$ intruder, for $^{99}$Zr $7/2^-$ 
intruder, and for $^{101,103}$Zr it is $5/2^-$ intruder 
(more details will be published elsewhere).

The analysis above is further stressed in 
\cref{fig:spe_nd_boson}, depicting the resulting 
SqPEs from the BCS calculation (panel a), and and the 
different occupations of the lowest positive-parity state, 
$J^+_\text{low}$: the SqP occupation 
\eqref{eq:prob_ibfm_spe} (panel b), $n_d$-U(5) 
boson symmetry occupation \eqref{eq:prob_ibfm_nd} (panel c) 
and configuration content \eqref{eq:prob_ibfm_bos} (panel 
d).

For $^{93,95}$Zr, panel (b) shows a dominant $\nu1d_{5/2}$ 
orbit, panel (c) shows a dominant $n_d=0$ component, and 
panel (d) shows a dominant normal component. 
This indicates a weak coupling scenario of a normal wave 
function with good U(5) boson symmetry coupled to a 
$\nu1d_{5/2}$ orbital. A similar pattern is observed for 
$^{97,99}$Zr, but with the $\nu2s_{1/2}$ orbital instead.
In contrast, for $^{101,103}$Zr, panel (b) shows a mixture 
of $\nu2s_{1/2}$, $\nu1d_{3/2}$ and $\nu0g_{7/2}$ orbitals 
within the intruder part of the wave function, panel (c) 
shows boson U(5)-$n_d$ occupations indicating a deformed 
intruder state, and panel (d) shows almost no normal 
occupation in the wave function, indicating an almost pure 
intruder state. 

The increase in $j$-mixture observed in panel (b) 
corresponds clearly to the lowering of the SqPEs shown in 
(a) of \cref{fig:spe_nd_boson}.
The evolution observed in panels (c) and (d), from 
spherical to deformed states and from normal to intruder 
configurations, indicates the presence of the two 
quantum phase transitions (IQPTs) similar to those seen in 
the even-even Zr and odd-mass Nb chains 
\cite{Gavrielov2024}.

To further examine the calculated wave functions, it is 
essential to compare them with electromagnetic transition 
rates.
For $E2$ transitions in \cref{eq:Tem1}, the boson effective 
charges $e^\text{(A)}, e^\text{(B)}$ are taken from 
\cite{Gavrielov2022}. The fermion effective 
charge, $e_f = 3.258~\sqrt{\text{W.u.}}$, is determined 
from a fit to the spectroscopic quadrupole moment of 
$^{95}$Zr, $Q_{5/2^+_1} = 0.22(2)~eb$.

The fitting procedure yields $Q_{3/2^+_1} = 
0.27~eb$ for $^{101}$Zr, compared to the experimental 
value of $Q=0.81(6)~eb$. The experimental $E2$ transitions 
(calculation in parentheses) are: 
$^{93}$Zr: $B(E2;3/2^+_1\to5/2^+_1) = 6.6(23)$~W.u. (1.1), 
$^{97}$Zr: $B(E2;7/2^+_1\to3/2^+_1) = 1.55(5)$~W.u. (4.80), 
$^{99}$Zr: $B(E2;7/2^+_1\to3/2^+_1) = 1.33(5)$~W.u. (1.91), 
and $^{101}$Zr: $B(E2;7/2^+_1\to3/2^+_1) > 130$~W.u. (82). 
The calculated values are reasonably close to the 
experimental ones and indicate the onset of deformation at 
$^{101}$Zr, as evidenced by the strong increase in 
transition strength, with a possible need for more 
deformation.

\begin{figure}[tb!]
\centering
\includegraphics[width=1\linewidth]{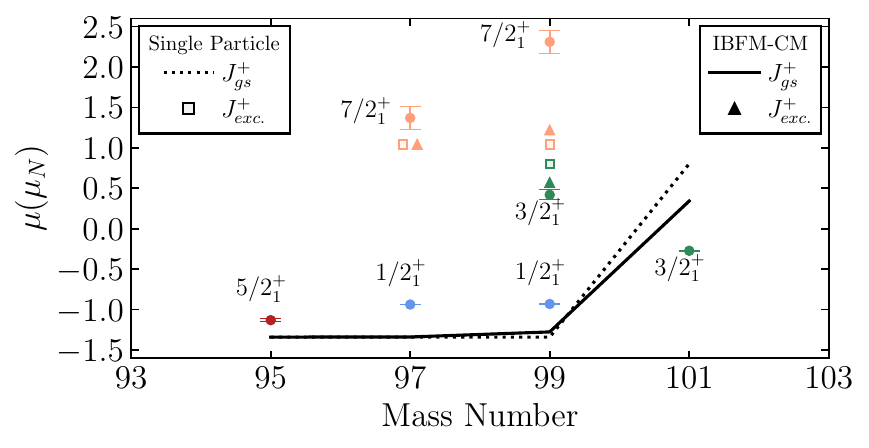}
\caption{Evolution of magnetic moments $\mu$ in units of 
$\mu_N$, along the odd Zr chain. Symbols denote 
experimental data for the ground state and excited states 
when available, taken from \cite{Stone2005}. Each $J$ value 
has a different color. The solid (dashed) line and black 
triangles (squares) denote calculated results, based on 
\cref{eq:tm1_b,eq:tm1_f} (single particle estimation). See 
text for more details.\label{fig:mag_mom}}
\end{figure}

Magnetic moments, which are strongly influenced by the 
single particle structure, provide further indicative 
observables. For $M1$ transitions operator, \cref{eq:Tem1}, 
the same values for $g^\text{(A)}, g^\text{(B)}, \tilde 
g^\text{(A)}, \tilde g^\text{(B)}$ is 
used as in \cite{Gavrielov2022c}. 
\cref{fig:mag_mom} compares the calculated magnetic 
moments with the experimental ones, as well as the single 
particle estimations \cite{TalmiBook1993}.
For $^\text{95--99}$Zr, there is a good agreement between 
the calculation and experimental data for the ground state. 
Notably, the IBFM-CM result also coincides with the single 
particle estimations \cite{TalmiBook1993}, further 
supporting the single particle character of the ground 
state. This interpretation extends to the $7/2^+_1$ state 
of $^{97}$Zr, which exhibits an almost pure $\nu0g_{7/2}$ 
occupation and spherical normal configuration in its 
wave function. 

However, deviations from the single particle 
interpretation are observed for the excited states of 
$^{99}$Zr and the ground state of $^{101}$Zr. 
For the $3/2^+_1$ and $7/2^+_1$ states in these isotopes, 
better agreement is achieved with the IBFM-CM calculation. 
The reason is that these states become more mixed in their 
orbitals, losing their single quasi-particle character due 
to the onset of deformation and configuration mixing. In 
$^{99}$Zr, the $3/2^+_1$ and $7/2^+_1$ are strongly mixed 
in configuration and $n_d$ occupations. For the $7/2^+_1$, 
there is still some deviation for the \mbox{IBFM-CM}, which 
might suggest more deformation is needed in the $^{98}$Zr 
core, as was suggested in \cite{Karayonchev2020}. For 
$^{101}$Zr, the $3/2^+_1$ state is pure intruder deformed, 
with mixed orbitals, which explains the deviation 
from the single particle value.

A closer examination of $^{99}$Zr is intriguing, as it lies 
nearest to the critical points of the shape evolution 
and configuration crossing QPTs between $^{98}$Zr and 
$^{100}$Zr \cite{Gavrielov2022}. Recently, an IBFM 
calculation with a single configuration was presented in 
\cite{Spagnoletti2019a, Boulay2020} for $^{99}$Zr. In 
\cite{Boulay2020}, the authors compared the $7/2^+_1$ state 
in $^{99}$Zr to that in $^{97}$Zr, noting that both states 
are isomers.
However, for the $7/2^+_1$ state, the magnetic moment for 
$^{99}$Zr, $\mu_{7/2^+1} = 2.31~(14)~\mu_N$, deviates from 
the single particle prediction observed in $^{97}$Zr. This 
led the authors in \cite{Boulay2020} to suggest the 
occurrence of a \mbox{type II} shell evolution, causing the 
\textit{single particle} $\nu2s{1/2}$ orbital to become 
lower than the $\nu1d_{5/2}$.

While Ref.~\cite{Boulay2020} successfully reproduced this 
magnetic moment value, the calculated energies did not 
align well with experimental data. This discrepancy was 
later questioned in Ref.~\cite{Garrett2021}, leading the 
authors to propose an IBFM configuration mixing model for 
improvement.
In \cite{Garrett2021}, it was argued that there is no need 
for a change in the single particle energies, as suggested 
by previous works \cite{Lhersonneau1990,Brant1998}. 
Furthermore, as was suggested in \cite{Togashi2016}, only 
the resulting \textit{effective} single particle energies 
are modified, leading to the \mbox{type~II} shell evolution 
between the spin orbit partners $\nu0g_{7/2}$ and 
$\pi0g_{9/2}$.

This work confirms that lowering the $\nu2s_{1/2}$ single 
particle orbital is unnecessary in $^{99}$Zr. Instead, the 
resulting single \textit{quasi}-particle energies account 
for the $1/2^+_1$ state as a normal single quasi-particle 
state of the $\nu2s_{1/2}$ orbital, while the $3/2^+_1$ and 
$7/2^+_1$ are mixed normal-intruder deformed states.
Furthermore, the intruder character of the $7/2^+_1$ state 
in $^{99}$Zr mainly results from coupling with the intruder 
$2^+_1$ state of $^{98}$Zr core (unlike the less 
conventional choice of $^{100}$Zr boson core with a 
hole in \cite{Boulay2020}). Its isomeric nature is 
attributed to the dominant single quasi-particle component 
$\nu0g_{7/2}$ in the normal-intruder components (total 
$v^2_{j=7/2^+_1} = 0.86$), while for the $3/2^+_1$ state, 
it is the $\nu1d_{3/2}$ orbital ($v^2_{j=3/2^+_1} = 0.73$), 
which are weakly connected via $\hat T_\text{f}(E2)$ in 
\cref{eq:Tem1}. 
In $^{97}$Zr, the isomeric nature arises from the 
single particle character of the $2^+_1$ state of $^{96}$Zr.

In summary, a comprehensive analysis of the odd 
$^\text{93--103}$Zr isotopes has been conducted using the 
interacting boson-fermion model with configuration mixing. 
This approach employed boson, fermion, and Bose-Fermi 
Hamiltonians with mixed configurations, focusing on 
positive-parity states. The good agreement with 
experimental data on energies, $E2$ transitions, and 
quadrupole and magnetic moments underscores the 
manifestation of shape, configuration, and single 
quasi-particle evolutions that influence the spectrum.
The special case of $^{99}$Zr was further examined in light 
of recent studies \cite{Spagnoletti2019a, Boulay2020} 
suggesting a type II shell evolution in the 
\textit{single}-particle $\nu2s_{1/2}$ orbit. It was 
demonstrated that while type II shell evolution occurs in 
the effective single particle orbits, as suggested in 
\cite{Togashi2016}, it does not occur in the single 
particle orbits and does not lower the $\nu2s_{1/2}$ orbit. 
Instead, the spectrum is shown to result from the interplay 
of shape, configuration, and single quasi-particle 
evolutions, offering a resolution to the recent debate 
\cite{Garrett2021, Boulay2021}. 
This work may inspire further experimental and theoretical 
efforts to identify and differentiate shape, configuration, 
and single particle effects in the structural evolution of 
atomic nuclei. As demonstrated here, these effects are 
particularly pronounced in odd-mass nuclei.

\begin{acknowledgments}
The author acknowledges support from the European Union's 
Horizon 2020 research and innovation program under the 
Marie Sk\l{}odowska-Curie grant agreement No. 101107805.
\end{acknowledgments}

\bibliography{refs.bib}

\end{document}